\documentclass[10pt,twocolumn,letterpaper]{article}

\usepackage{iccv}
\usepackage{times}
\usepackage{epsfig}
\usepackage{graphicx}
\usepackage{amsmath}
\usepackage{amssymb}

\usepackage{subfig}
\usepackage{makecell}
\usepackage{sidecap}

\usepackage{booktabs}
\usepackage{multirow}
\usepackage{multicol}
\usepackage{makecell}

\usepackage{amsmath,amsfonts,bm}

\def\eqref#1{equation~(\ref{#1})}

\def\1{\bm{1}}

\DeclareMathAlphabet{\mathsfit}{\encodingdefault}{\sfdefault}{m}{sl}
\SetMathAlphabet{\mathsfit}{bold}{\encodingdefault}{\sfdefault}{bx}{n}

\def\gL{{\mathcal{L}}}

\DeclareMathOperator*{\argmax}{arg\,max}
\DeclareMathOperator*{\argmin}{arg\,min}

\usepackage[pagebackref=true,breaklinks=true,letterpaper=true,colorlinks,bookmarks=false]{hyperref}

\def\iccvPaperID{5508}

\ificcvfinal\fi

\begin{document}

\title{Revisiting Adversarial Robustness Distillation: Robust Soft Labels \\Make Student Better}

\author{Bojia Zi$^{1,2}$\footnotemark[1], Shihao Zhao$^{1,2}$\footnotemark[1], Xingjun Ma$^{3}$\footnotemark[2], Yu-Gang Jiang$^{1,2}$\footnotemark[2]\\
\normalsize$^{1}$Shanghai Key Lab of Intelligent Information Processing, School of Computer Science, Fudan Univeristy\\
\normalsize$^{2}$Shanghai Collaborative Innovation Center on Intelligent Visual Computing\\
\normalsize$^{3}$School of Information Technology, Deakin University, Geelong, Australia
}

\maketitle
\ificcvfinal\thispagestyle{empty}\fi

\renewcommand{\thefootnote}{\fnsymbol{footnote}}
\footnotetext[1]{Equal contribution: Bojia Zi(bjzi19@fudan.edu.cn) and Shihao Zhao(shzhao19@fudan.edu.cn)}
\footnotetext[2]{Correspondence to Xingjun Ma (daniel.ma@deakin.edu.au) and Yu-Gang Jiang (ygj@fudan.edu.cn)}

\begin{abstract}
\vspace{-0.1in}
\label{abstract}

Adversarial training is one effective approach for training robust deep neural networks against adversarial attacks. While being able to bring reliable robustness, adversarial training (AT) methods in general favor high capacity models, i.e., the larger the model the better the robustness. This tends to limit their effectiveness on small models, which are more preferable in scenarios where storage or computing resources are very limited (e.g., mobile devices).
In this paper, we leverage the concept of knowledge distillation to improve the robustness of small models by distilling from adversarially trained large models.
We first revisit several state-of-the-art AT methods from a distillation perspective and identify one common technique that can lead to improved robustness: the use of robust soft labels --  predictions of a robust model. Following this observation, we propose a novel adversarial robustness distillation method called Robust Soft Label Adversarial Distillation (RSLAD) to train robust small student models.
RSLAD fully exploits the robust soft labels produced by a robust (adversarially-trained) large teacher model to guide the student's learning on both natural and adversarial examples in all loss terms.
We empirically demonstrate the effectiveness of our RSLAD approach over existing adversarial training and distillation methods in improving the robustness of small models against state-of-the-art attacks including the AutoAttack. We also provide a set of understandings on our RSLAD and the importance of robust soft labels for adversarial robustness distillation. Code: \href{https://github.com/zibojia/RSLAD}{https://github.com/zibojia/RSLAD}.

\end{abstract}
\vspace{-0.1in}

\section{Introduction}
\label{introduction}

Deep Neural Networks (DNNs) have become the standard models for solving complex real-world learning problems, such as image classification \cite{krizhevsky2017imagenet,he2015deep}, speech recognition \cite{wang2017residual} and natural language processing \cite{vaswani2017attention}. However, studies have shown that DNNs are vulnerable to adversarial attacks \cite{szegedy2014intriguing,goodfellow2015explaining}, where imperceptible adversarial perturbations on the input can easily subvert the model's prediction. This raises security concerns on the deployment of DNNs in safety-critical scenarios such as autonomous driving \cite{eykholt2018robust, chen20203d,duan2020adversarial} and medical diagnosis \cite{ma2021understanding}.

Different types of methods have been proposed to defend DNNs against adversarial attacks \cite{jia2019comdefend,ma2018characterizing,madry2019deep,kurakin2017adversarial,zhang2019theoretically,wang2019improving}, amongst which adversarial training (AT) has been found to be the most effective approach \cite{athalye2018obfuscated,croce2020reliable}. 
AT can be regarded as a type of data augmentation technique that crafts adversarial versions of the natural examples for model training. AT is normally formulated as a min-max optimization problem with the inner maximization generates adversarial examples while the outer minimization optimizes the model's parameters on the adversarial examples generated during the inner maximization \cite{madry2019deep,zhang2019theoretically,wang2019convergence}.

While being able to bring reliable robustness, AT methods have several drawbacks that may limit their effectiveness in certain application scenarios. Arguably, the most notable drawback is its hunger for high capacity models, i.e., the larger the model the better the robustness \cite{wu2021wider, uesato2019labels,rice2020overfitting,gowal2020uncovering}. However, there are scenarios where small and lightweight models are more preferable than large models. One example is the deployment of small DNNs in devices with limited memory and computational power such as smart phones and autonomous vehicles \cite{sandler2019mobilenetv2}.
This has motivated the use of knowledge distillation along with AT to boost the robustness of small DNNs by distilling from robust large models \cite{Goldblum_2020,bai2020feature,chen2021robust,zhu2021reliable}, a process known as Adversarial Robustness Distillation (ARD).

In this paper, we build upon previous works in both AT and ARD, and investigate the key element that can boost the robustness of small DNNs via distillation. We compare the loss functions adopted by several state-of-the-art AT methods and identify one common technique behind the improved robustness: the use of predictions of an adversarially trained model. We denote this type of supervision as \emph{Robust Soft Labels} (RSLs). Compared to the original hard labels, RSLs can better represent the robust behaviors of the teacher model, providing more robust information to guide the student's learning.
This observation motivates us to design a new ARD method to fully exploit the power of RSLs in boosting the robustness of small student models.

In summary, our main contributions are:
\begin{itemize}
    \item We identify that the implicit distillation process existing in adversarial training methods is a useful function for promoting robustness and the use of \emph{robust soft labels} can lead to improved robustness.
    \item We propose a novel adversarial robustness distillation method called Robust Soft Label Adversarial Distillation (RSLAD), which applies \emph{robust soft labels} to replace hard labels in all of its supervision loss terms.
    \item We empirically verify the effectiveness of RSLAD in improving the robustness of small DNNs against state-of-the-art attacks. We also provide a comprehensive understanding of our RSLAD and the importance of robust soft labels for robustness distillation. 
\end{itemize}

\section{Related Work}
\label{related works}

\subsection{Adversarial Attack}
Given a DNN model with known parameters, adversarial examples (or attacks) can be crafted by Fast Gradient Sign Method (FGSM) \cite{goodfellow2015explaining}, Projected Gradient Descent (PGD) \cite{madry2019deep}, Carlini and Wagner (CW) attack \cite{carlini2017evaluating} and a number of other methods. Several recent attacks were developed to produce more reliable adversarial robustness evaluation of defense models. These methods were designed to effectively avoid subtle gradient masking or obfuscating effects in improperly defended models. The AutoAttack (AA) \cite{croce2020reliable} is an ensemble of four attacking methods including Auto-PGD (APGD), Difference of Logits Ratio (DLR) attack, FAB-Attack \cite{croce2020minimally} and the black-box Square Attack \cite{andriushchenko2020square}. The AA ensemble is arguably the most powerful attack to date.

\subsection{Adversarial Training}
\label{conventional adversarial training}
Adversarial training is known as the most effective approach to defend adversarial examples. Recently, a number of understandings~\cite{ma2018characterizing,DBLP:conf/nips/IlyasSTETM19,Engstrom2019AdversarialRA,zhang2019interpreting,zhao2021deep}
and methods~\cite{madry2019deep,zhang2019theoretically,wang2019improving,xie2019feature,xie2021smooth,qin2019adversarial,gowal2021uncovering,guo2020nas,bai2020improving} have been put forward in this area.
Adversarial training can be formulated as the following min-max optimization problem:
\begin{equation}
\label{eq_adv}
\begin{aligned}
    &\underset{\textup{Outer minimization } }{\underbrace{\argmin \limits_{\theta}\gL_{\textup{min}}(f(x', \theta), y)}} \\
    \textup{where} \quad & x' =
    \underset{\textup{Inner maximization}}{\underbrace{\argmax\limits_{\left\|x'-x\right\|_p\leq\epsilon} \gL_{\textup{max}}(f(x', \theta), y)}}
\end{aligned}
\end{equation}
where $f$ is a DNN model with parameters $\theta$, $x'$ is the adversarial example of natural example $x$ within bounded $L_{p}$ distance $\epsilon$, $\gL_{\textup{min}}$ is the loss for the outer minimization, $\gL_{\textup{max}}$ is the loss for the inner maximization. The most commonly adopted $L_{p}$ norm is the $L_{\infty}$ norm.
In Standard Adversarial Training (SAT) \cite{madry2019deep}, the two losses $\gL_{\textup{min}}$ and $\gL_{\textup{max}}$ are set to the same loss, i.e., the most commonly used Cross Entropy (CE) loss. And the inner maximization problem is solved by the PGD attack. For simplicity, we omit the $\theta$ from the loss functions for the rest of this paper.

A body of work has been proposed to further improve the effectiveness of SAT. This includes the use of wider and larger models \cite{wu2021wider}, additional unlabeled data \cite{carmon2019unlabeled}, domain adaptation (natural domain versus adversarial domain) \cite{song2019improving}, theoretically-principled trade-off between robustness and accuracy (known as TRADES) via the use of Kullback–Leibler (KL) divergence loss for $\gL_{\textup{max}}$, the emphasis of misclassified examples via Misclassification-Aware adveRsarial Training (MART) \cite{wang2019improving}, channel-wise activation suppressing (CAS) \cite{bai2020improving} and adversarial weight perturbation \cite{wu2020adversarial}. In general, elements that have been found in these works that can contribute to robustness include large models, more data, and the use of KL loss for the inner maximization.

AT methods are not perfect. One notable drawback of existing AT methods is that the smaller the model the poorer the robust performance \cite{gowal2020uncovering}. It is generally hard to improve the robustness of small models like ResNet-18 \cite{he2015deep} and MobileNetV2 \cite{sandler2019mobilenetv2}, though many of the above AT methods can bring considerable robustness improvements to large models such as WideResNet-34-10 \cite{zhang2019theoretically,wang2019improving} and WideResNet-70-16 \cite{gowal2020uncovering}. 
This tends to limit their effectiveness in scenarios where storage or computational resources are limited, such as mobile devices, autonomous vehicles and drones.
In this paper, we leverage knowledge distillation techniques to improve the robustness of small models and improve existing adversarial robustness distillation methods.

\subsection{Knowledge Distillation}
\label{knowledge distillation}
Knowledge distillation (KD) is one well-known method for deep neural network compression that distills the knowledge of a large DNN into a small, lightweight student DNN \cite{hinton2015distilling}. Given a well-trained teacher network $T$, KD trains the student network $S$ by solving the following optimization problem:
\begin{equation}
\underset{\theta_{S}}{\textup{argmin}} (1-\alpha)\gL(S(x),y) + \alpha\tau^{2}\textup{KL}(S^{\tau}(x),T^{\tau}(x)),
\end{equation}
where $\textup{KL}$ is the Kullback-Leibler divergence, $\tau$ is a temperature constant added to the softmax operation, $\gL$ is the classification loss of the student network with CE is a common choice.
KD has been extended in different ways \cite{romero2015fitnets,zagoruyko2017paying,li2020reskd,zhang2019teacher} to a variety of learning tasks, such as noisy label learning \cite{xie2020selftraining,zhang2020distilling}, AI security \cite{Goldblum_2020,bai2020feature,li2021neural} and natural language processing \cite{nakashole-flauger-2017-knowledge,sun2020mobilebert,lu2020twinbert}.
Notably, a branch called self-distillation has attracted considerable attention in recent years \cite{kim2020selfknowledge,zhang2019teacher,xu_2019_self_distillation}. Unlike traditional KD methods, self-distillation teaches a student network by itself rather than a separate teacher network.

KD has been applied along with adversarial training to boost the robustness of a student network with an adversarially pre-trained teacher network. The teacher can be a larger model with better robustness\cite{Goldblum_2020} (e.g. ARD) or share the same architecture with the student \cite{zhu2021reliable} (e.g. IAD).
It has been shown that ARD and IAD can produce student networks that are more robust than trained from scratch, indicating that robust features learned by the teacher network can also be distilled \cite{bai2020feature}. In this paper, we will build upon these works and propose a more effective adversarial robustness distillation method to improve the robustness of small student networks.

\begin{table*}[!htb]
\begin{center}
\vspace{-0.075in}
\caption{A unified view of 6 defense methods from the perspective of knowledge distillation. $\gL_{\textup{min}}$ is the loss function for the outer minimization while $\gL_{\textup{max}}$ is the loss function for the inner maximization. S and T represent the student and the teacher network respectively. $\lambda$ in TRADES, MART and $\alpha$ in ARD, RSLAD are parameters balancing the two loss terms in $\gL_{\textup{min}}$. $\tau$ is a temperature constant added to the softmax operation. $\beta$ in IAD is a hyper-parameter to sharpen the prediction.}
\label{tab_loss_compare}
\setlength{\tabcolsep}{2.5mm}{
\begin{tabular}{c|c|c|c}
\toprule
\textbf{Method} & \textbf{$\gL_{\textup{min}}$} & \textbf{$\gL_{\textup{max}}$} & Student/Teacher \\
\midrule
SAT & $\textup{CE}(f(x'),y)$ & $\textup{CE}(f(x'),y)$ & - \\ \hline
TRADES & $\textup{CE}(f(x),y) + \lambda\textup{KL}(f(x'),f(x))$ & $\textup{KL}(f(x'),f(x))$ & S: $f(\cdot)$; T: $f(\cdot)$\\ \hline
MART & $\textup{BCE}(f(x'),y)+\lambda  (1-f_{y}(x))\textup{KL}(f(x'),f(x))$ & $\textup{CE}(f(x'),y)$ & S: $f(\cdot)$; T: $f(\cdot)$\\ \hline
ARD & $(1-\alpha)\textup{CE}(S^{\tau}(x),y)+\alpha\tau^{2}\textup{KL}(S^{\tau}(x'),T^{\tau}(x))$ & $\textup{CE}(S(x'),y)$ & S: $S(\cdot)$; T: $T(\cdot)$\\ \hline
IAD & $T_{y}(x')^\beta\textup{KL}(S^{\tau}(x'),T^{\tau}(x))+(1-T_{y}(x')^\beta)\textup{KL}(S^{\tau}(x'),S^{\tau}(x))$ & $\textup{CE}(S(x'),y)$ & S: $S(\cdot)$; T: $T(\cdot)$\\ \hline
\textbf{RSLAD} (ours) & $(1-\alpha)\textup{KL}(S(x),T(x))+\alpha\textup{KL}(S(x'),T(x))$ & $\textup{KL}(S(x'),T(x))$ & S: $S(\cdot)$; T: $T(\cdot)$\\ \bottomrule
\end{tabular}
}
\vspace{-0.12in}
\end{center}
\end{table*}

\section{Proposed Distillation Method}
\label{rsad}

In this section, we revisit state-of-the-art AT and adversarial robustness distillation methods from the perspective of KD, and identify the importance of using robust soft labels for improving robustness. We then introduce our RSLAD method inspired by robust soft labels.

\subsection{A Distillation View of Adversarial Training}
\label{revisiting adversarial training}

Following the adversarial training framework defined in \eqref{eq_adv}, we summarize, in Table \ref{tab_loss_compare}, the loss functions and the student and teacher networks used in 4 state-of-the-art AT methods (i.e., SAT \cite{madry2019deep}, TRADES \cite{zhang2019theoretically} and MART \cite{wang2019improving}) and two adversarial robustness distillation methods (i.e., ARD \cite{Goldblum_2020} and IAD \cite{zhu2021reliable}). 
Compared to SAT which simply adopts the original hard label to supervise the learning, TRADES utilizes the natural predictions of the model via the KL term and gains significant robustness improvement \cite{zhang2019theoretically}. From this perspective, TRADES is a self-distillation process where the teacher network is the student itself. MART \cite{wang2019improving} is also a self-distillation process but with a focus on the low probability examples via the (1-$f_y(x)$) weighting scheme on the KL term.
In ARD, a more powerful teacher instead of the student itself is used to supervise the learning. The robustness is constantly improved from SAT's no distillation, TRADES/MART's self-distillation to ARD's full distillation \cite{Goldblum_2020}, as we will also show in Section \ref{experiments}.
IAD \cite{zhu2021reliable} is also an adversarial distillation method, which makes the distillation process more reliable by using the knowledge of both the teacher and the student networks.
In this view, we believe that knowledge distillation implicitly or explicitly adopted in these methods contributes significantly to their success.

Another key difference between SAT and other methods mentioned above is that the latter exploit the teacher network's natural predictions in both of their outer and inner optimization processes, via the KL term.
The predictions of a robust teacher model can be considered as a type of \emph{Robust Soft Labels} (RSLs). 
Previous works (and also our experiments in Section \ref{experiments}) have shown that TRADES and its variants can bring considerable robustness improvement to SAT. From a distillation point of view, this robustness improvement comes from the use of RSLs, contrasting the use of original hard labels $y$.
On the other hand, adversarial robustness distillation is to make the student as similar to the robust teacher as possible. 
Compared to the original hard labels, RSLs define the full robust behavior of the teacher network, thus convey more robust knowledge learned by the teacher to the student. In Section \ref{experiments}, we will empirically show that RSLs are indeed more beneficial to robustness than the original hard labels or other forms of non-robust soft labels. ARD has a KL term in its outer minimization loss, however, its other loss terms use the original hard labels $y$.
IAD uses the KL terms in its two outer minimization loss terms, but the inner maximization loss still uses the hard labels, leaving space for improvement.

\begin{figure*}[!htb]
\centering
\includegraphics[width=.875\linewidth]{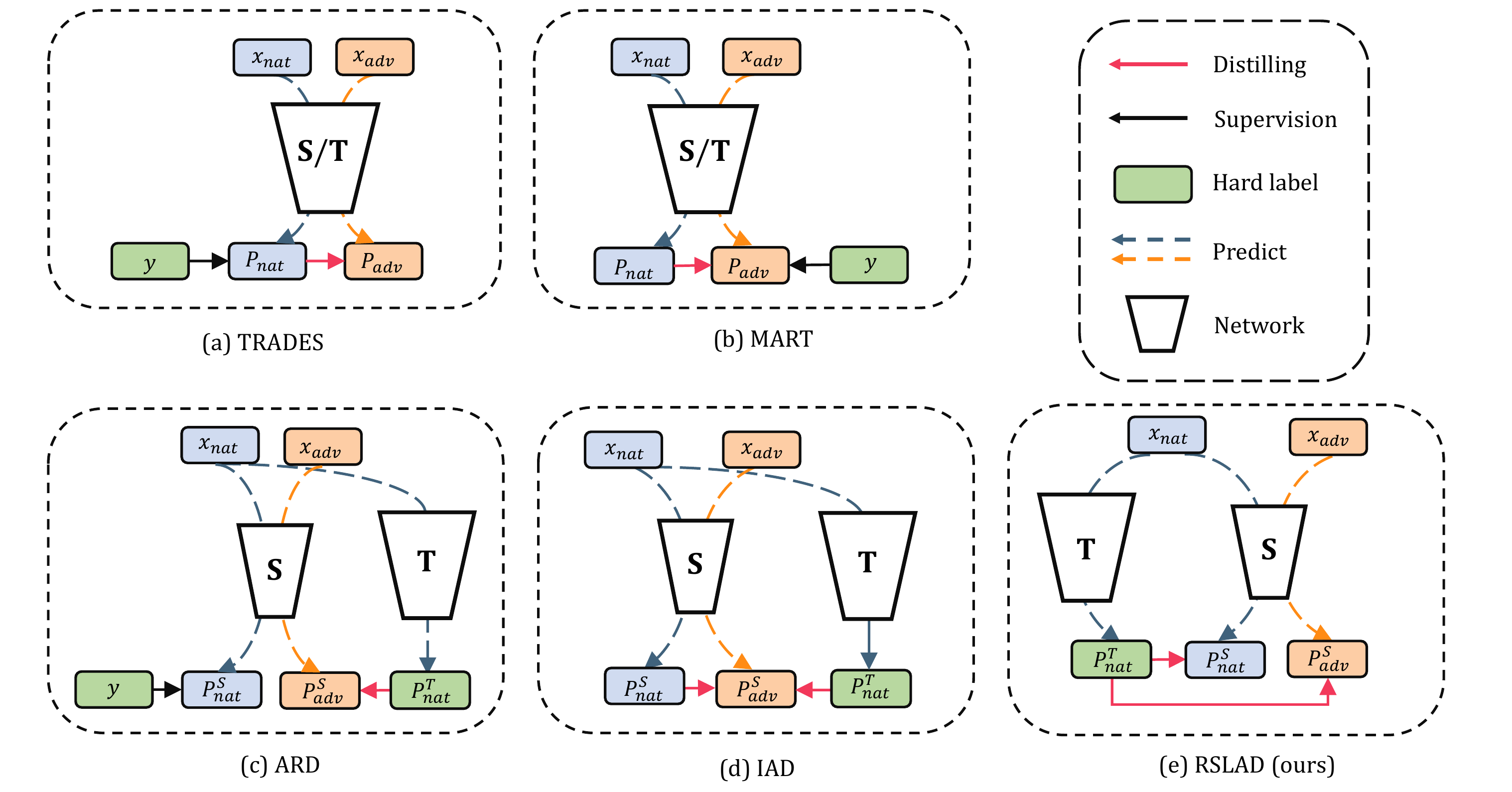}
\caption{An overview of the proposed \textbf{RSLAD} framework, in comparison with 4 existing methods including \textbf{TRADES}, \textbf{MART}, \textbf{ARD} and \textbf{IAD}. Black solid arrows represent training with hard labels $y$; yellow and blue dashed arrows represent predicting process for natural and adversarial examples respectively; red solid arrows represent distillation using \emph{robust soft labels}. $\textbf{S}$ and $\textbf{T}$ represent the student and the teacher network respectively. $P_{nat}$ and $P_{adv}$ are the predictions of the model for the natural examples $x_{nat}$ and adversarial examples $x_{adv}$. Note that no hard labels $y$ are used in our RSLAD.}
\label{fig_framework}
\vspace{-0.15in}
\end{figure*}

\subsection{Robust Soft Label Adversarial Distillation}
\label{boosting adversarial distillation}

The proposed Robust Soft Label Adversarial Distillation (RSLAD) framework is illustrated in Figure \ref{fig_framework}, including a comparison with four existing methods (i.e., TRADES, MART, ARD and IAD). 
The key difference of our RSLAD to existing methods lies in the use of RSLs produced by the large teacher network to supervise the student's training on both natural and adversarial examples in all loss terms. 
The original hard labels $y$ are absent in our RSLAD.

As the student network in RSLAD is still trained using AT, it also has the inner maximization and the outer minimization processes. 
To bring RSLs into its full play, we apply RSLs in both of the two processes. The loss functions used by our RSLAD are summarized in the last row of Table \ref{tab_loss_compare}.
Note that, in our RSLAD, the temperature constant commonly exists in distillation methods is fixed to $\tau=1$ as we find it is no longer necessary when RSLs are used. Same as TRADES, MART, ARD and IAD, we use the natural RSLs (i.e. the predictions of a robust model for natural examples) as the soft label to supervise the model training.

\begin{table*}[h]
\setlength{\abovecaptionskip}{0.2cm}
\setlength{\belowcaptionskip}{-0.7cm}
\begin{center} 
\caption{Robustness of the teacher networks used in our experiments.}
\vspace{-0.05in}
\label{tab_teacher}
\setlength{\tabcolsep}{1mm}{
\begin{tabular}{cc|ccccccc}
\toprule
Dataset & Teacher & Clean & FGSM & PGD$_{\textup{SAT}}$ & PGD$_{\textup{TRADES}}$ & CW$_{\infty}$ & AA \\
\midrule
CIFAR-10 & WideResNet-34-10 &84.92\% & 60.87\% & 55.33\% & 56.61\% & 53.98\% & 53.08\% \\
CIFAR-100 & WideResNet-34-10 &57.16\% & 33.58\% & 30.61\% & 31.34\% & 27.74\% & 26.78\% \\
CIFAR-100 & WideResNet-70-16 &60.86\% & 35.68\% & 33.56\% & 33.99\% & 42.15\% & 30.03\% \\
\bottomrule
\end{tabular}
}
\vspace{-0.15in}
\end{center}
\end{table*}

\begin{table*}[h]\small
\begin{center}
\caption{White-box robustness results on CIFAR-10 dataset. MN-V2 and RN-18 are abbreviations of MobileNetV2 and ResNet-18 respectively. The maximum adversarial perturbation is $\epsilon=8/255$. The best results are \textbf{blodfaced}.}
\vspace{-0.08in}
\label{tab_eval_cifar10}
\setlength{\tabcolsep}{0.55mm}{
\begin{tabular}{c|c|cccccc|cccccc}
\toprule
\multirow{2}{*}{Model}&\multirow{2}{*}{Method} & \multicolumn{6}{c|}{\textbf{Best Checkpoint}} & \multicolumn{6}{c}{\textbf{Last Checkpoint}} \\
 && Clean & FGSM & PGD$_{\textup{SAT}}$ & PGD$_{\textup{TRADES}}$ & CW$_{\infty}$ & AA & Clean & FGSM & PGD$_{\textup{SAT}}$ & PGD$_{\textup{TRADES}}$ & CW$_{\infty}$ & AA\\
\midrule
&Natural & \textbf{94.65\%} & 19.26\% & 0.0\% & 0.0\% & 0.0\% & 0.0\% & \textbf{94.65\%} & 19.26\% & 0.0\% & 0.0\% & 0.0\% & 0.0\%\\
&SAT & 83.38\% & 56.41\% & 49.11\% & 51.11\% & 48.67\% & 45.83\% & 84.44\% & 55.37\% & 46.22\% & 48.72\% & 47.14\% & 43.64\%\\
RN-18&TRADES & 81.93\% & 57.49\% & 52.66\% & 53.68\% &50.58\%& 49.23\% & 82.20\% & 57.86\% & 52.30\% & 53.66\% & 50.69\% & 49.27\%\\
&ARD & 83.93\% & 59.31\% & 52.05\% & 54.20\% & 51.22\% & 49.19\% & 84.23\% & 59.33\% & 51.52\% & 53.74\% & 51.24\% & 48.90\%\\
&IAD & 83.24\% & 58.60\% & 52.21\% & 54.18\% & 51.25\% & 49.10\% & 83.90\% & 58.95\% & 51.35\% & 53.15\% & 50.52\% & 48.48\%\\
&\textbf{RSLAD} & 83.38\% & \textbf{60.01\%} & \textbf{54.24\%} & \textbf{55.94\%} & \textbf{53.30\%} & \textbf{51.49\%} & 83.33\% & \textbf{59.90\%} & \textbf{54.14\%} & \textbf{55.61\%} & \textbf{53.22\%} & \textbf{51.32\%}\\
\bottomrule
&Natural & \textbf{92.95\%} & 14.47\% & 0.0\% & 0.0\% & 0.0\% & 0.0\% & \textbf{92.78\%} & 14.59\% & 0.0\% & 0.0\% & 0.0\% & 0.0\%\\
&SAT & 82.48\% & 56.44\% & 50.10\% & 51.74\% & 49.33\% & 46.32\% & 82.89\% & 56.43\% & 49.71\% & 51.48\% & 49.07\% & 45.92\%\\
MN-V2&TRADES & 80.57\% & 56.05\% & 51.06\% & 52.36\% & 49.36\% & 47.17\% & 80.57\% & 56.05\% & 51.06\% & 52.36\% & 49.36\% & 47.17\%\\
&ARD & 83.20\% & 58.06\% & 50.86\% & 52.87\% & 50.39\% & 48.34\% & 83.42\% & 57.94\% & 50.63\% & 52.44\% & 50.09\% & 48.01\%\\
&IAD & 81.91\% & 57.00\% & 51.88\% & 53.23\% & 50.45\% & 48.40\% & 83.49\% & 57.44\% & 49.77\% & 51.85\% & 49.41\% & 46.98\%\\
&\textbf{RSLAD} & 83.40\% & \textbf{59.06\%} & \textbf{53.16\%} & \textbf{54.78\%} & \textbf{51.91\%} & \textbf{50.17\%} & 83.11\% & \textbf{59.08\%} & \textbf{53.04\%} & \textbf{54.50\%} & \textbf{51.60\%} & \textbf{49.90\%}\\
\bottomrule
\end{tabular}
}
\vspace{-0.18in}
\end{center}
\end{table*}

\begin{table*}[h]\small

\begin{center}
\caption{White-box robustness results on CIFAR-100 dataset. MN-V2 and RN-18 are abbreviations of MobileNetV2 and ResNet-18 respectively. The maximum adversarial perturbation is $\epsilon=8/255$. The best results are \textbf{blodfaced}.}
\vspace{-0.1in}
\label{tab_eval_cifar100}
\setlength{\tabcolsep}{0.6mm}{
\begin{tabular}{c|c|cccccc|cccccc}
\toprule
\multirow{2}{*}{Model}&\multirow{2}{*}{Method} & \multicolumn{6}{c|}{\textbf{Best Checkpoint}} & \multicolumn{6}{c}{\textbf{Last Checkpoint}} \\
 && Clean & FGSM & PGD$_{\textup{SAT}}$ & PGD$_{\textup{TRADES}}$ & CW$_{\infty}$ & AA & Clean & FGSM & PGD$_{\textup{SAT}}$ & PGD$_{\textup{TRADES}}$ & CW$_{\infty}$ & AA\\
\midrule
&Natural & \textbf{75.55\%} & 9.48\% & 0.0\% & 0.0\% & 0.0\% & 0.0\% & \textbf{75.39\%} &  9.57\%& 0.0\% & 0.0\% & 0.0\% & 0.0\%\\
&SAT & 57.46\% & 28.56\% & 24.07\% & 25.39\% & 23.68\% & 21.79\% & 57.51\% & 26.41\% &  21.7\% &  23.30\%& 22.15\% & 20.44\%\\
RN-18&TRADES & 55.23\% & 30.48\% & 27.79\% & 28.53\% & 25.06\% & 23.94\% & 54.62\% & 30.06\% & 27.35\% & 28.00\% & 24.34\%& 23.42\%\\
&ARD & 60.64\% & 33.41\% & 29.16\% & 30.30\% & 27.85\% & 25.65\% & 60.86\% & 32.64\% & 28.15\% & 29.34\% & 26.79\% & 24.74\%\\
&IAD & 57.66\% & 33.26\% & 29.59\% & 30.58\% & 27.37\% & 25.12\% & 58.82\% & 33.22\% & 28.50\% & 29.97\% & 26.79\% & 24.79\%\\
&\textbf{RSLAD} & 57.74\% &\textbf{34.20\%} & \textbf{31.08\%} & \textbf{31.90\%} & \textbf{28.34\%} & \textbf{26.70\%} & 57.82\% & \textbf{34.06\%} & \textbf{30.68\%} & \textbf{31.57\%} & \textbf{28.16\%} & \textbf{26.34\%}\\
\bottomrule
&Natural & \textbf{74.58\%} & 7.19\% & 0.0\% & 0.0\% & 0.0\% & 0.0\% & \textbf{74.58\%} & 7.19\% & 0.0\% & 0.0\% & 0.0\% & 0.0\%\\
&SAT & 56.85\% & 31.95\% & 28.33\% & 29.50\% & 26.85\% & 24.71\% & 58.50\% & 32.05\% & 27.80\% & 28.88\% & 26.74\% & 24.31\%\\
MN-V2&TRADES & 56.20\% & 31.37\% & 29.21\% & 29.83\% & 25.06\% & 24.16\% & 56.56\% & 31.35\% & 28.85\% & 29.38\% & 25.00\% & 24.04\%\\
&ARD & 59.83\% & 33.05\% & 29.13\% & 30.26\% & 27.86\% & 25.53\% & 61.66\% & 32.98\% & 27.74\% & 29.33\% & 26.77\% & 24.34\%\\
&IAD & 56.14\% & 32.81\% & 29.81\% & 30.73\% & 27.99\% & 25.74\% & 58.07\% & 32.61\% & 27.55\% & 28.81\% & 26.24\% & 23.72\%\\
&\textbf{RSLAD} & 58.97\% & \textbf{34.03\%} & \textbf{30.40\%} & \textbf{31.36\%} & \textbf{28.22\%} & \textbf{26.12\%} & 58.76\% & \textbf{34.02\%}& \textbf{30.17\%} & \textbf{31.14\%} & \textbf{28.10\%} & \textbf{26.31\%}\\
\bottomrule
\end{tabular}
}
\vspace{-0.2in}
\end{center}
\end{table*}

\begin{table}[h]\footnotesize
\begin{center}
\caption{Black-box robustness results on CIFAR-10 dataset. The maximum adversarial perturbation is $\epsilon=8/255$. The best results are \textbf{blodfaced}.}
\vspace{-0.08in}
\label{tab_black_box}
\setlength{\tabcolsep}{0.6mm}{
\begin{tabular}{c|ccc|ccc}
\toprule
\multirow{2}{*}{Method} & \multicolumn{3}{c|}{\textbf{ResNet-18}} & \multicolumn{3}{c}{\textbf{MobileNetV2}} \\
&PGD-20 & CW$_{\infty}$ & Square &PGD-20& CW$_{\infty}$&Square \\
\midrule
SAT  & 60.84\%&60.52\%&54.27\%&60.46\%&59.83\%&53.94\%\\
TRADES  & 62.20\%&61.75\%&55.13\%&60.90\%&60.23\%&53.46\%\\
ARD  & 63.49\%&63.05\%&56.89\%&62.13\%&61.85\%&55.60\%\\
IAD  & 62.78\%&62.26\%&56.62\%&61.57\%&61.25\%&55.45\%\\
\textbf{RSLAD}  & \textbf{64.11\%}&\textbf{63.84\%}&\textbf{57.90\%}&\textbf{63.30\%}&\textbf{63.20\%}&\textbf{56.70\%}\\
\bottomrule
\end{tabular}
}
\vspace{-0.3in}
\end{center}
\end{table}

\begin{table*}
\begin{center}
\caption{Ablation study with ResNet-18 student network distilled using variants of our RSLAD and ARD \cite{Goldblum_2020}. \textbf{ARD-300}: ARD training under our RSLAD setting (i.e. 300 epochs); \textbf{ARD$_{\textup{min}}$}: the outer maximization part of ARD; \textbf{ARD$_{\textup{max}}$}: the inner minimization part of ARD; \textbf{RSLAD$_{\textup{min}}$}: the outer minimization part of our RSLAD; \textbf{RSLAD$_{\textup{max}}$}: the inner maximization part of our RSLAD.}
\vspace{-0.075in}
\label{tab_ablation}
\setlength{\tabcolsep}{3.2mm}{
\begin{tabular}{cccccccc}
\toprule
Distillation Method & Clean & FGSM & PGD$_{\textup{SAT}}$ &  PGD$_{\textup{TRADES}}$ & $\textup{CW}_{\infty}$ & AA \\
\midrule
ARD-300 & 84.40\% & 59.81\% & 52.36\% & 54.49\% & 51.58\% & 49.70\% \\
ARD$_{\textup{min}}$+RSLAD$_{\textup{max}}$& \textbf{84.70\%} & \textbf{60.77\%} & 52.99\% & 54.84\% & 52.09\% & 50.35\% \\
RSLAD$_{\textup{min}}$+ARD$_{\textup{max}}$& 84.44\% & 59.89\% & 53.10\% & 55.01\% & 52.15\% & 49.94\%\\
\textbf{RSLAD} & 83.38\% & 60.01\% & \textbf{54.24\%} & \textbf{55.94\%} & \textbf{53.30\%} & \textbf{51.49\%} \\
\bottomrule
\end{tabular}
}
\vspace{-0.2in}
\end{center}
\end{table*}

The overall optimization framework of our RSLAD is defined as following:
\begin{equation}
\label{eq_rsad}
\begin{split}
    \argmin \limits_{\theta_S} (1-&\alpha) \textup{KL}(S(x),T(x)) +\alpha \textup{KL}(S(x'),T(x)) \\
    \textup{where}& \quad x' = \argmax\limits_{\left\|x'-x\right\|_p\leq\epsilon} \textup{KL}(S(x),T(x))
\end{split}
\end{equation}
where $S(x)$ and $T(x)$ are the abbreviations for $S(x,\theta_{S})$ and $T(x,\theta_{T})$, respectively.
Since the RSLs produced by the adversarially trained teacher network $T(x)$ are also used to supervise the clean training part of the student's outer minimization, here we replace the commonly used CE loss by KL divergence to formulate the degree of distributional difference between the two models' output probabilities.

The goal of RSLAD is to learn a small student network that is as robust as an adversarially pre-trained teacher network, which is also to retain as much as possible the teacher's knowledge and robustness.
We note that the commonly used hard labels in adversarial training can lose information learned by the teacher network to some extent, due to the fact that binarizing the teacher's output probabilities into hard labels tends to lose its true distribution. However, not all soft labels are robust. We will empirically show that smooth labels produced by label smoothing or soft labels produced by naturally trained non-robust models cannot improve robustness.

\section{Experiments}
\label{experiments}
We first describe the experimental setting, then evaluate the white-box robustness of 4 baseline defense methods and our RSLAD. We also conduct an ablation study, visualize the attention map learned by different methods, compare 3 types of soft labels, and explore how to choose a better teacher network. 

\subsection{Experimental Settings}
\label{experiment settings}
We conduct our experiments on two benchmark datasets including CIFAR-10 and CIFAR-100 \cite{cifar10}, and consider 5 baseline methods: SAT \cite{madry2019deep}, TRADES \cite{zhang2019theoretically}, ARD \cite{Goldblum_2020}, IAD \cite{zhu2021reliable} and natural training.

\noindent\textbf{Student and Teacher Networks.}
We consider two student networks including ResNet-18 \cite{he2015deep} and MobileNetV2 \cite{sandler2019mobilenetv2}, and two teacher networks including WideResNet-34-10 \cite{zagoruyko2017wide} for CIFAR-10 and WideResNet-70-16 \cite{gowal2020uncovering} for CIFAR-100.
The CIFAR-10 teacher WideResNet-34-10 is trained using TRADES, while for CIFAR-100, we use the WideResNet-70-16 model provided by Gowal et al. \cite{gowal2020uncovering}.

\noindent\textbf{Training Setting.}
We train the networks using Stochastic Gradient Descent (SGD) optimizer with initial learning rate 0.1, momentum 0.9 and weight decay 2e-4. We set batch size to 128. For our RSLAD, we set the total number of training epochs to 300, and the learning rate is divided by 10 at the 215th, 260th and 285th epoch. A 10 step PGD (PGD-10) with random start size  0.001, step size 2/255 is used to solve the inner maximization of our RSLAD.
For baseline methods SAT, TRADES and ARD, we strictly follow their original settings. 
IAD uses the same structure for the teacher and student networks. Here, we reproduce their method by using a more powerful teacher to fit our settings.
Training perturbation is bounded to the $L_{\infty}$ norm $\epsilon=8/255$ for both datasets.
For natural training, we train the networks for 100 epochs on clean images with standard data augmentations and the learning rate is divided by 10 at the 75th and 90th epochs.

\noindent\textbf{Evaluation Attacks.}
After training, we evaluate the model against 5 adversarial attacks: FGSM, $\textup{PGD}_{\textup{SAT}}$, $\textup{PGD}_{\textup{TRADES}}$, $\textup{CW}_{\infty}$(optimized by PGD) and AutoAttack(AA).
The $\textup{PGD}_{\textup{SAT}}$ attack is the original attack proposed in Madry et al. \cite{madry2019deep}, while $\textup{PGD}_{\textup{TRADES}}$ is the one used in Zhang et al. \cite{zhang2019theoretically}. They both are PGD attacks but differ in their hyper-parameters (e.g. step size). We consider these two attacks separately following Carmon et al. \cite{carmon2019unlabeled}.
Note these attacks are commonly used adversarial attacks in adversarial robustness evaluation. 
Maximum perturbation used for evaluation is also set to $\epsilon=8/255$ for both datasets. The perturbation steps of $\textup{PGD}_{\textup{SAT}}$,
$\textup{PGD}_{\textup{TRADES}}$ and $\textup{CW}_{\infty}$ are all 20.
The robustness of the teacher models against the 5 attacks are reported in Table \ref{tab_teacher}, indicating the maximum robustness the student model can get. Besides the white-box evaluation, we also conduct a black-box evaluation which will be described later.

\subsection{Adversarial Robustness Evaluation}
\label{the performance of rsad}

\noindent\textbf{White-box Robustness.}\label{robustness_evaluation}
The white-box robustness of our RSLAD and other baseline methods are reported in Table \ref{tab_eval_cifar10} for CIFAR-10 and Table \ref{tab_eval_cifar100} for CIFAR-100.
Following previous works, we report the results at both the best checkpoint and the last checkpoint.
The best checkpoint of naturally training (i.e., showing as `Natural' in both Tables) is selected based on the performance on clean test examples, and the best checkpoints of SAT, TRADES, ARD, IAD, and our RSLAD are selected based on their robustness against the $\textup{PGD}_{\textup{TRADES}}$ attack.

\begin{figure*}[t]
\centering
\includegraphics[width=0.95\linewidth]{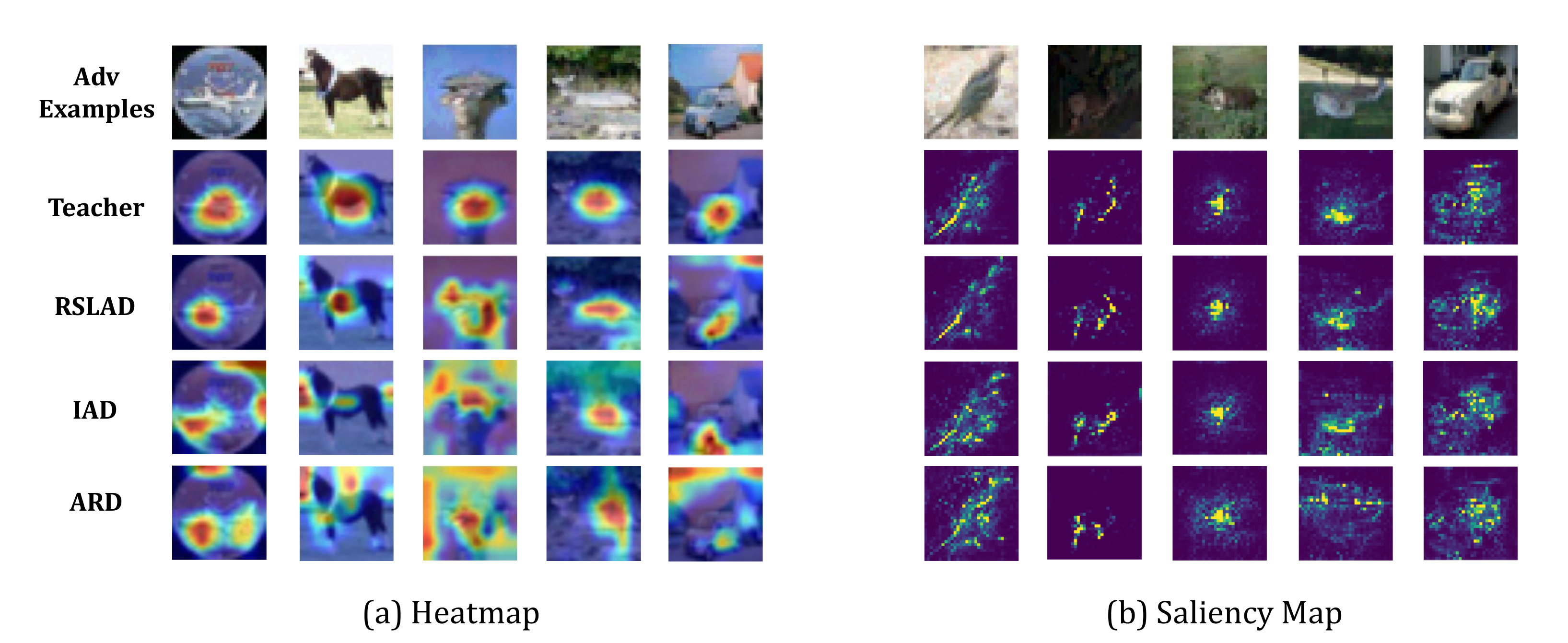}
\vspace{-0.1in}
\caption{Attention and saliency maps on adversarial examples. \textbf{Teacher:} WideResNet-34-10 trained by TRADES; \textbf{ARD:} the ResNet-18 student trained using ARD with the \textbf{Teacher} network; \textbf{RSLAD:} the ResNet-18 student trained using our RSLAD with the \textbf{Teacher} network. Heatmaps are generated by Grad-Cam\cite{Selvaraju_2019} while saliency maps are generated by \cite{smilkov2017smoothgrad}.}
\vspace{-0.2in}
\label{fig_vis}
\end{figure*}

As shown in Table \ref{tab_eval_cifar10} and Table \ref{tab_eval_cifar100}, our RSLAD method demonstrates the state-of-the-art robustness on both CIFAR-10 and CIFAR-100 against all 5 attacks at either the best or the last checkpoints.
For ResNet-18, RSLAD improves the robustness by 1.74\% and 1.32\% on CIFAR-10 and CIFAR-100 respectively, compared to previous SOTA under $\textup{PGD}_{\textup{TRADES}}$ attack.
For MobileNetV2, RSLAD brings 1.55\% and 0.63\% improvements against the $\textup{PGD}_{\textup{TRADES}}$ attack.
The improvements are more pronounced against the AutoAttack, which is the most powerful attack to date.
Particularly, our RSLAD outperforms ARD by even 2.30\% for the ResNet-18 student on CIFAR-10.
This verifies that our RSLAD is more stable and robust in training robust small DNNs than all the baseline methods.
We also observe that, under all settings, TRADES holds clear advantage over SAT, but can still be largely outperformed by distillation methods (i.e., ARD and our RSLAD).

\noindent\textbf{Black-box Robustness.}
Here, we evaluate the black-box robustness of our RSLAD, SAT, TRADES, ARD and IAD. We test both the transfer attack and query-based attack. This experiment is conducted on CIFAR-10 dataset.
For transfer attack, we craft the test adversarial examples using 20 step PGD (PGD-20) and CW$_\infty$ on an adversarially pre-trained ResNet-50 surrogate model. The maximum perturbation is also set to $8/255$. For query-based attack, we use one strong and query-efficient attack, i.e., the Square attack, to attack the models.
We evaluate both the transfer attack and query-based attacks on the best checkpoints of the two student models (i.e., ResNet-18 and MobileNetV2). The results are presented in Table \ref{tab_black_box}.
As can be observed, our RSLAD surpasses all the 4 baseline methods against all 3 black-box attacks, demonstrating the superiority of our robust soft label distillation approach.
The general trend across different types of defense methods is consistent with that in the white-box setting: for robustifying small DNNs, TRADES is better than SAT while distillation methods are better than TRADES.

\subsection{A Comprehensive Understanding of RSLAD}
\label{ablation study}

\noindent\textbf{Ablation of RSLAD.}
To better understand the impact of each component of our RSLAD to robustness, we conduct a set of ablation studies with the existing distillation method ARD on CIFAR-10 with the ResNet-18 student network (the teacher is the same WideResNet-34-10 network as used in the above experiments).
We replace the inner maximization and outer minimization losses used by ARD by the ones used in our RSLAD, then test the robustness of the trained student network.
We also run an experiment with ARD under our RSLAD setting for 300 epochs (it was 200 epochs in the original paper).
The ablation results are reported in Table \ref{tab_ablation}. Compared to ARD, there is a certain improvement when either the inner loss or the outer loss of our RSLAD is used.
The best robustness is achieved when both losses in ARD are switched to our RSLAD losses.
This confirms the importance of each component of RSLAD, and the robust soft labels used in these components.
We also find that the outer maximization has more impact on the overall robustness than the inner minimization: replacing the inner part of ARD by RSLAD leads to a more robust student than the outer part.
An additional comparison between RSLAD and the baselines trained for 300 epochs can be found in Appendix \ref{sec:appendix-d}.

\noindent\textbf{Attention Maps Learned by RSLAD.}
Here we use attention maps and saliency maps to visually inspect the similarity of the knowledge learned by the student to that of the teacher network. Given the same adversarial examples, higher similarity indicates more successful distillation and better aligned robustness to the teacher model.
We take the ResNet-18 student distilled from the WideResNet-34-10 teacher on CIFAR-10 dataset as an example, and visualize the attention maps (generated by Grad-CAM \cite{Selvaraju_2019}) and saliency maps (generated by \cite{smilkov2017smoothgrad}) in Figure \ref{fig_vis}.
As can be observed, the attention maps of the student trained using our RSLAD are noticeably more similar to that of the teacher's than baseline methods ARD and IAD. This indicates that the student trained by our RSLAD can indeed mimic the teacher better and has gained more robust knowledge from the teacher. 
A parameter analysis of our RSLAD can be found in the appendix.

\subsection{Further Explorations}
\label{soft label comparison}

\begin{figure}[!htp]
\centering
\includegraphics[width=0.95\linewidth]{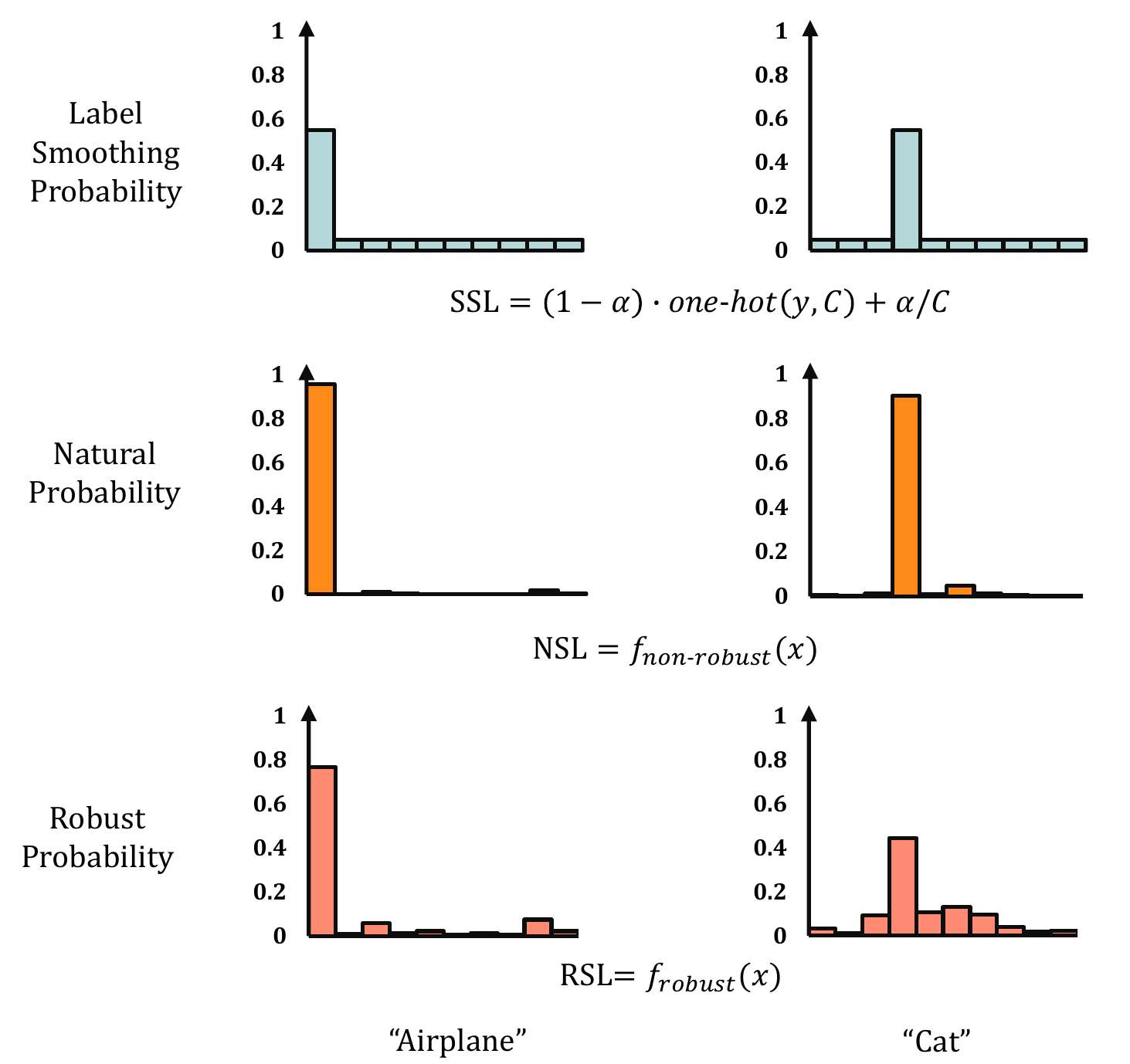}
\caption{Probability distributions of 3 types of soft labels. $f_{robust}$ represents the robust model, which is the adversarial trained model, $f_{non-robust}$ represents the non-robust model, which is the standard trained model. $C$ represent the number of class in dataset, $one$-$hot(\cdot,\cdot)$ stands for the function that convert the label $y$ to one-hot vector, $\alpha$ is a parameter which adjust the max number of the vector.}
\label{fig_soft_label}
\vspace{-0.15in}
\end{figure}

\begin{table}[h]\footnotesize
\begin{center}
\caption{White-box robustness of ResNet-18 student trained using our RSLAD with three types of soft labels (i.e., SSL, NSL and RSL). The best results are \textbf{boldfaced}.}
\label{tab_soft_label}
\setlength{\tabcolsep}{0.55mm}{
\begin{tabular}{c|ccc|ccc}
\toprule
\multirow{2}{*}{Soft Labels} & \multicolumn{3}{c|}{\textbf{Best Checkpoint}} & \multicolumn{3}{c}{\textbf{Last Checkpoint}} \\
 &Clean & $\textup{PGD}_{\textup{TRADES}}$ & AA & Clean & $\textup{PGD}_{\textup{TRADES}}$ & AA\\
\midrule
SSL  & \textbf{85.67\%} & 53.12\% & 47.88\% & \textbf{85.26\%} & 49.70\% & 43.92\%\\
NSL& 85.02\% & 47.12\% & 42.87\% & 84.99\% & 46.69\% & 42.08\%\\
\textbf{RSL} & 83.38\% & \textbf{55.94\%} & \textbf{51.49\%} & 83.33\% & \textbf{55.61\%} & \textbf{51.32\%} \\
\bottomrule
\end{tabular}
}
\vspace{-0.2in}
\end{center}

\end{table}

\noindent\textbf{Different Types of Soft Labels.}
Here, we compare three types of soft labels: 1) smooth soft labels (SSLs) crafted by label smoothing \cite{szegedy2015rethinking}; 2) natural soft labels (NSLs) produced by a naturally trained teacher model; and 3) robust soft labels (RSLs) produced by an adversarially trained robust teacher model. 
This experiment is conducted with ResNet-18 student and  WideResNet-34-10 teacher on CIFAR-10 dataset, with our RSLAD. The probability distributions of the three types of soft labels for two example CIFAR-10 classes (i.e., `Airplane' and `Cat') are plotted in Figure \ref{fig_soft_label}.
Different to RSLs, SSLs implement a fixed smoothing transformation to the original hard labels, while NSL probabilities are more concentrated around the ground truth label. 
The white-box robustness of the student network trained using our RSLAD with these 3 types of soft labels are shown in Table \ref{tab_soft_label}. One key observation is that the robustness drops drastically when non-robust labels including SSLs or NSLs are used in place of robust labels. This means that soft labels are not all beneficial to robustness, and non-robust labels especially the NSLs produced by non-robust models can significantly harm robustness distillation.

\begin{figure}[!htp]
\centering
\includegraphics[width=0.9\linewidth]{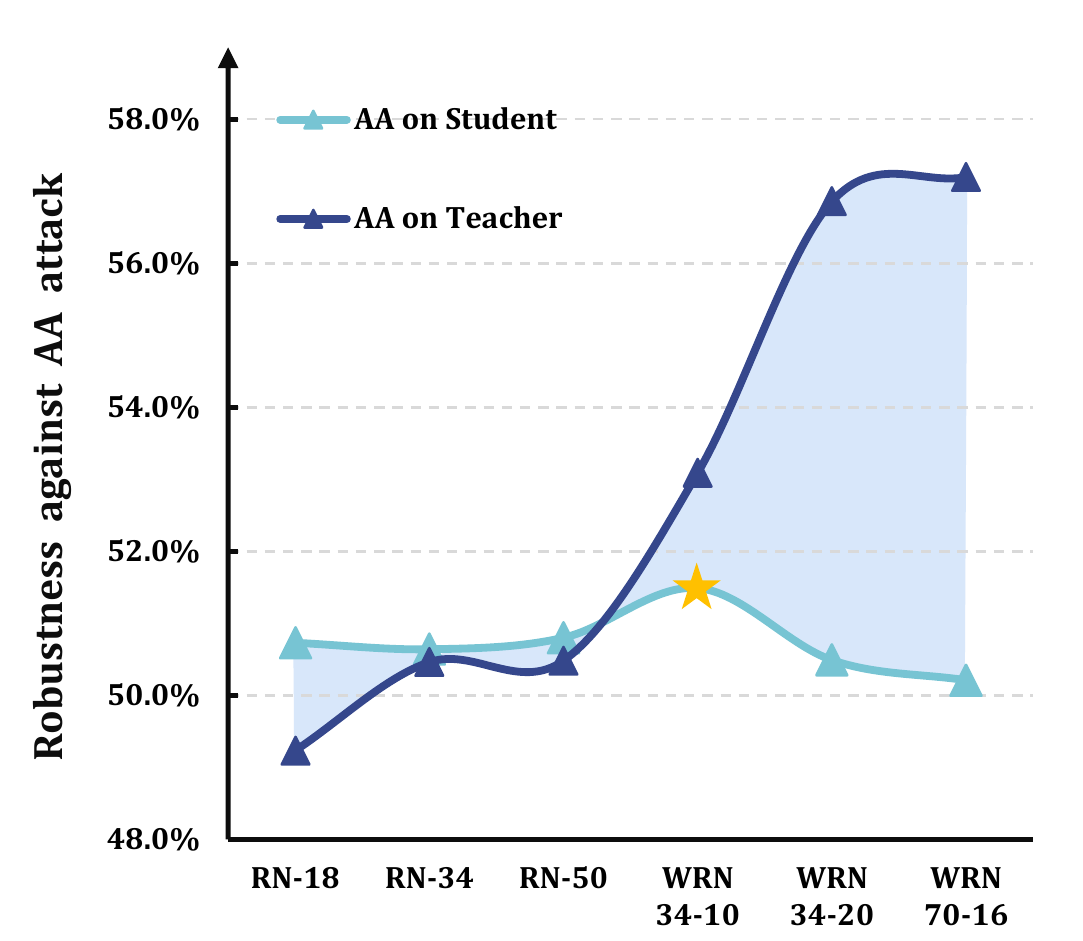}
\vspace{-0.075in}
\caption{Robustness against AA attack of ResNet-18 (RN-18) students trained using our RSLAD with 6 different teachers. RN: ResNet; WRN: WideResNet. The RN-18, RN-34, RN-50, WRN-34-10 teachers are trained using TRADES, while the rest teacher models are from Gowal et al. \cite{gowal2020uncovering}. This experiment is done on CIFAR-10 dataset.
}
\label{fig_better_teacher}
\vspace{-0.18in}
\end{figure}

\noindent\textbf{How to Choose a Good Teacher?}
Here, we provide some empirical understandings on the impact of the teacher on the robustness of the student. We conduct this experiment on CIFAR-10 with the ResNet-18 student network and investigate its robustness when distilled using our RSLAD from 6 different teacher networks: ResNet-18, ResNet-34, ResNet-50, WideResNet-34-10, WideResNet-34-20 and WideResNet-70-16. The results are plotted in Figure \ref{fig_better_teacher}. Surprisingly, we find that the student's robustness does not increase monotonically with that of the teacher's, instead, it first rises then drops. We call this phenomenon \emph{robust saturation}. When the teacher network becomes too complex for the student to learn, the robustness of the student tends to drop. As shown in the figure, the robustness gap between the student and the teacher increases when the complexity of the teacher network goes beyond WideResNet-34-10.
Interestingly, the student's robustness can surpass that of the teacher's when the teacher is smaller than WideResNet-34-10, especially when the teacher has the same architecture (i.e., ResNet-18) as the student. We call this phenomenon the \emph{robust underfitting} of adversarial training methods, where robustness can be improved by training the model the second time while using the model trained the first time as a teacher. The robust underfitting region is where distillation can help boost the robustness. The best robustness of the ResNet-18 student is achieved when the WideResNet-34-10 ($\sim$4.5$\times$ larger than ResNet-18) teacher is used. These results indicate that choosing a moderately large teacher model can lead to the maximum robustness gain in adversarial robustness distillation.

\section{Conclusion}
\label{conclusion}

In this paper, we investigated the problem of training small robust models via knowledge distillation.
We revisited several state-of-the-art adversarial training and robustness distillation methods from the perspective of distillation. By comparing their loss functions, we identified the importance of robust soft labels (RSLs) for improved robustness.
Following this view, we proposed a novel adversarial robustness distillation method named Roust Soft Label Adversarial Distillation (RSLAD) to fully exploit the advantage of RSLs. The advantage of RSLAD over existing adversarial training and distillation methods were empirically verified on two benchmark datasets under both the white-box and the black-box settings. We also provided several insightful understandings of our RSLAD, different types of soft labels, and more importantly, the interplay between the teacher and student networks. Our work can help build adversarially robust lightweight deep learning models.

\section*{ACKNOWLEDGEMENT}
This work was supported in part by National Natural Science Foundation of China (\#62032006) and STCSM (\#20511101000).

{\small
\bibliographystyle{ieee_fullname}
\bibliography{egbib}
}

\clearpage
\appendix
\section{Hyper-parameter Selection}
\label{hyper-parameters selection}
\begin{figure}[!htp]
\centering
\includegraphics[width=0.9\linewidth]{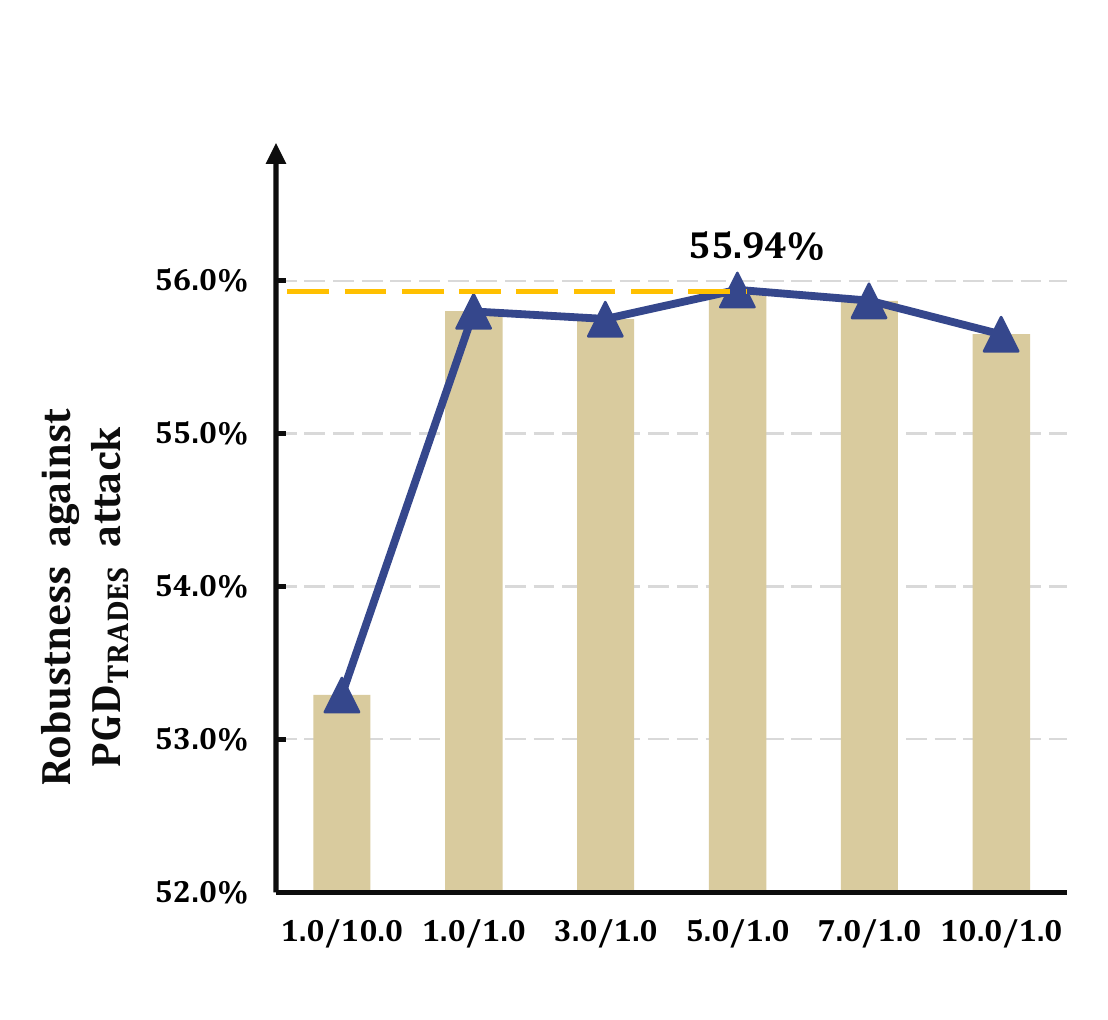}
\caption{Robustness on CIFAR-10 for ResNet-18 student distilled by our RSLAD with WideResNet-34-10 teacher, under different hyper-parameters $\alpha$. 
\textbf{X-axis:} ratio of the adversarial term to the natural term: $k=\frac{\alpha}{(1-\alpha)}$; \textbf{Y-axis:} robustness against $\textup{PGD}_{\textup{TRADES}}$ on CIFAR-10 test set.
}
\label{fig_hyper_p}
\end{figure}

In this section, we explore the impact of the hyper-parameter $\alpha$ in \eqref{eq_rsad}. We apply RSLAD to train and distill ResNet-18 students from the the WideResNet-34-10 teacher on CIFAR-10 using different $\alpha$. 
We show the robust accuracy against PGD$_\textup{TRADES}$ with respect to $k = \frac{\alpha}{(1-\alpha)}$, which is the ratio of the adversarial loss term to the natural loss term. We report the robustness results at the best checkpoints in Figure \ref{fig_hyper_p}.
It can be observed that robustness rises rapidly with the increase of the ratio $k$ and reaches a plateau after $k=1.0$. When the ratio becomes larger than 1, the robustness fluctuates slightly around 55.9\% and achieves the best at $k=5.0/1.0$.

\begin{table*}[!ht]
\begin{center}
\caption{Robustness of the teacher networks used in our experiments on CIFAR-10 dataset. The maximum perturbation is $\epsilon=8/255$. The best results are \textbf{blodfaced}.}
\label{tab_teachers}
\setlength{\tabcolsep}{0.75mm}{
\begin{tabular}{c|cccccc}\hline
Teacher&Clean&FGSM&PGD$_{\textup{SAT}}$&PGD$_{\textup{TRADES}}$&CW$_{\infty}$&AA \\ \hline
ResNet-18&81.93\%&57.49\%&52.66\%&53.68\%&50.58\%&49.23\% \\ 
ResNet-34&83.38\%&59.54\%&53.70\%&55.25\%&52.39\%&50.46\% \\ 
ResNet-50&84.25\%&60.27\%&53.71\%&55.31\%&52.47\%&50.47\% \\ 
WideResNet-34-10&84.92\%&60.87\%&55.33\%&56.61\%&53.98\%&53.08\% \\ 
WideResNet-34-20&\textbf{85.64\%}&\textbf{64.29\%}&\textbf{59.86\%}&\textbf{60.82\%}&58.04\%&56.86\% \\ 
WideResNet-70-16&85.29\%&64.20\%&59.66\%&60.46\%&\textbf{58.60\%}&\textbf{57.20\%} \\ \hline

\end{tabular}
}
\end{center}
\end{table*}

\section{Learning From Different Teachers}
In Section \ref{soft label comparison}, we have demonstrated how to choose a good teacher network and showed the impact of the teacher on the student's robustness.
Here, we show a more complete robustness results of the student network against all 5 attacks mentioned in Section \ref{experiment settings}.
We report results at both the best and the last checkpoints in Table \ref{tab_eval_resnet18}. The teachers' robustness is shown in Table \ref{tab_teachers}.
We can confirm the phenomenon of \emph{robust saturation} and \emph{robust underfitting} according to more evaluation attacks. This indicates that a moderately large teacher network can be a better teacher than a overly large teacher network.

\begin{table*}[!htp]\small
\begin{center}
\caption{Robustness of ResNet-18 student distilled using our RSLAD with 6 different teacher networks. Both the best and last checkpoints are reported and the best results are \textbf{blodfaced}.}
\label{tab_eval_resnet18}
\setlength{\tabcolsep}{0.6mm}{
\begin{tabular}{c|c|cccccc|cccccc}
\toprule
\multirow{2}{*}{Student}&\multirow{2}{*}{Teacher} & \multicolumn{6}{c|}{\textbf{Best Checkpoint}} & \multicolumn{6}{c}{\textbf{Last Checkpoint}} \\
 && Clean & FGSM & PGD$_{\textup{SAT}}$ & PGD$_{\textup{TRADES}}$ & CW$_{\infty}$ & AA & Clean & FGSM & PGD$_{\textup{SAT}}$ & PGD$_{\textup{TRADES}}$ & CW$_{\infty}$ & AA\\
\midrule
&RN-18 & 81.14\% & 58.62\% & 53.92\% & 55.31\% & 52.08\% & 50.75\% & 81.43\% &  58.61\%& 53.63\% & 54.90\% & 51.90\% & 50.65\%\\
&RN-34 & 81.96\% & 59.00\% & 53.94\% & 55.19\% & 51.95\% & 50.64\% & 81.79\% & 59.12\% &  53.78\% & 55.10\%& 52.03\% & 50.61\%\\
RN-18&RN-50 & 82.19\% & 59.39\% & 54.09\% & 55.55\% & 52.24\% & 50.80\% & 82.46\% & 59.45\% & 53.57\% & 55.26\% & 51.93\%& 50.29\%\\
&WRN-34-10 & \textbf{83.38\%} & 60.01\% & \textbf{54.24\%} & \textbf{55.94\%} & \textbf{53.30\%} & \textbf{51.49\%} & \textbf{83.33\%} & 59.90\% & 54.14\% & \textbf{55.61\%} & \textbf{53.22\%} & \textbf{51.32\%}\\
&WRN-34-20 & 83.36\% & \textbf{60.11\%} & 54.18\% & 55.85\% & 51.58\% & 50.49\% & 83.25\% & \textbf{60.31\%} & \textbf{54.24\%} & 55.60\% & 51.86\% & 50.50\%\\
&WRN-70-16 & 82.96\% & 59.61\% & 53.63\% & 55.17\% & 51.82\% & 50.27\% & 82.99\% & 59.42\% & 53.36\% & 54.96\% & 51.64\% & 50.04\%\\
\bottomrule
\end{tabular}
}
\end{center}
\end{table*}

\section{RSLs of Natural or Adversarial Examples?}
RSLs are the outputs of a robust model, however, it can be on either the natural examples (natural RSLs $T(x)$) or the adversarial examples (adversarial RSLs $T(x^{\prime})$). Same as TRADES, MART, ARD and IAD, RSLAD utilizes the $T(x)$ as RSLs. 
But one may wonder whether $T(x^{\prime})$ is better than $T(x)$.
To answer this question, we replace the $T(x)$ used in our RSLAD loss terms (Equation 3) with $T(x^{\prime})$. This experiment is conducted with ResNet-18 student and WideResNet-34-10 teacher on CIFAR-10 dataset. 
We report the results at the best checkpoints in Table \ref{tab_rsl}. 

Looking at the first row of Table \ref{tab_rsl}, one can find that, when replacing $T(x^{\prime})$ with $T(x)$ in all loss terms, the robustness reaches 49.79\% against AA, which is slightly higher than that of the TRADES (49.27\%) (see Table 2), but still much lower than our RSLAD with $T(x)$.
Moving on to the second and third rows, one may notice that, when replacing more $T(x')$ with $T(x)$ in $\mathcal{L}_{max}$ or $\mathcal{L}_{min}$, both clean accuracy and robustness can be improved.
This clearly demonstrates the advantage of using RSLs of the natural examples, albeit RSLs of adversarial examples also help robustness.

\begin{table*}[!ht]\small
\begin{center}
\caption{Robustness of ResNet-18 student trained using different types of RSLs ($x$: clean examples; $x'$: adversarial examples) on CIFAR-10 dataset. The maximum perturbation is $\epsilon=8/255$. The best results are \textbf{blodfaced}.}
\label{tab_rsl}
\setlength{\tabcolsep}{0.75mm}{
\begin{tabular}{c|c|cccccc}\hline
$\mathcal{L}_{min}$&$\mathcal{L}_{max}$&Clean&FGSM&PGD$_{\textup{SAT}}$&PGD$_{\textup{TRADES}}$&CW$_{\infty}$&AA \\ \hline
$(1-\alpha)\textup{KL}(S(x),\textbf{T}(\textbf{x}^{\prime}))+\alpha\textup{KL}(S(x^{\prime}),\textbf{T}(\textbf{x}^{\prime}))$&$\textup{KL}(S(x^{\prime}),\textbf{T}(\textbf{x}^{\prime}))$&78.86\%&57.16\%&52.81\%&54.12\%&51.34\%&49.79\%\\ 
$(1-\alpha)\textup{KL}(S(x),T(x))+\alpha\textup{KL}(S(x^{\prime}),\textbf{T}(\textbf{x}^{\prime}))$&$\textup{KL}(S(x^{\prime}),\textbf{T}(\textbf{x}^{\prime}))$& 79.04\%&57.53\%&53.14\%&54.41\%&51.37\%&49.83\%\\ 
$(1-\alpha)\textup{KL}(S(x),T(x))+\alpha\textup{KL}(S(x^{\prime}),\textbf{T}(\textbf{x}^{\prime}))$&$\textup{KL}(S(x^{\prime}),T(x))$&82.95\%&59.81\%&54.13\%&55.91\%&53.06\%&51.26\% \\ 
$(1-\alpha)\textup{KL}(S(x),T(x))+\alpha\textup{KL}(S(x^{\prime}),T(x))$&$\textup{KL}(S(x^{\prime}),T(x))$&\textbf{83.38\%}&\textbf{60.01\%}&\textbf{54.24\%}&\textbf{55.94\%}&\textbf{53.30\%}&\textbf{51.49\%}\\ \hline
\end{tabular}
}
\end{center}
\end{table*}
\section{Training all baselines for 300 epochs}\label{sec:appendix-d}
Whilst the training epoch is 100 in SAT, TRADES and 200 in ARD, IAD, we train our RSLAD models for 300 epochs. For a fair comparison, here we also run the baseline methods for 300 epochs.
Table \ref{tab:300-epoch} shows the results of 300-epoch SAT, TRADES, ARD and IAD on CIFAR-10, following the setting in Section \ref{experiment settings}.
It shows that, although the robustness of \textbf{best} checkpoints has been slightly improved when training for 300 epochs using TRADES, ARD and IAD, their performances on the \textbf{last} checkpoints degrade, resulting in more overfitting. As expected, our RSLAD method achieves the best overall performance.

\begin{table*}[h]
\begin{center}
\setlength{\tabcolsep}{0.6mm}{
\caption{Robustness results of ResNet-18 on CIFAR-10. The best results are \textbf{boldfaced}. -300 means 300 epochs of training. }
\label{tab:300-epoch}
\scalebox{0.9}{
\begin{tabular}{c|cccccc|cccccc}
\toprule
\multirow{2}{*}{Method} & \multicolumn{6}{c|}{\textbf{Best Checkpoint}} & \multicolumn{6}{c}{\textbf{Last Checkpoint}} \\
 & Clean& FGSM& PGD$_{\textup{SAT}}$&PGD$_{\textup{TRADES}}$& CW$_\infty$&AA & Clean& FGSM&PGD$_{\textup{SAT}}$&PGD$_{\textup{TRADES}}$ &CW$_\infty$ &AA\\
\midrule
Natural &\textbf{94.65\%}&0.0\%&0.0\%&0.0\%&0.0\%&0.0\%&\textbf{94.65\%}&0.0\%&0.0\% & 0.0\%& 0.0\%& 0.0\%\\ \hline
SAT & 83.38\% & 56.41\%& 49.11\%&51.11\%&  48.67\%&45.83\% & 84.44\% &  55.37\%& 46.22\%&48.72\%& 47.14\%&43.64\% \\
SAT-300 & 83.96\%& 55.42\%& 47.04\%&49.21\%& 47.47\%&44.73\% & 84.29\%  &52.08\%& 42.42\%&44.69\%&  48.33\%&40.99\% \\ \hline
TRADES & 81.93\% &  57.49\%& 52.66\%&53.68\%& 50.58\%&49.23\% & 82.20\%& 57.86\%& 52.30\%&53.66\%&  50.69\%&49.27\% \\
TRADES-300 & 82.06\% &  57.97\%& 52.65\%&53.96\%& 50.91\%&49.50\% & 82.79\%  &57.50\%& 49.97\%&51.83\%&  49.51\%&47.59\% \\ \hline
ARD &83.93\% & 59.31\% & 52.05\% & 54.20\% & 51.22\% & 49.19\% & 84.23\% & 59.33\% & 51.52\% & 53.74\% & 51.24\% & 48.90\% \\
ARD-300 & 84.40\% & 59.81\% & 52.36\% & 54.49\% & 51.58\% & 49.70\%& 85.01\%& 55.42\%&51.59\%&53.59\% & 50.98\% &48.72\% \\ \hline
IAD & 83.24\% & 58.60\% & 52.21\% & 54.18\% & 51.25\% & 49.10\% & 83.90\% & 58.95\% & 51.35\% & 53.15\% & 50.52\% & 48.48\% \\
IAD-300 & 83.68\% &  59.20\%& 52.83\%& 54.58\%& 51.84\%& 49.54\%& 84.
35\%& 59.92\%&51.30\%&53.44\% & 50.61\% &48.60\% \\ \hline
RSLAD & 83.38\% & \textbf{60.01\%} & \textbf{54.24\%} & \textbf{55.94\%} & \textbf{53.30\%} & \textbf{51.49\%} & 83.33\% & \textbf{59.90\%} & \textbf{54.14\%} & \textbf{55.61\%} & \textbf{53.22\%} & \textbf{51.32\%} \\
\bottomrule
\end{tabular}
}}
\end{center}
\end{table*}

\section{Teacher models' diversity}\label{section:e}
Considering that we use a teacher trained by TRADES in most experiments in the main text, to figure out whether our method works with different kinds of teachers, we try to train the teacher model using SAT+AWP. AWP~\cite{wu2020adversarial} is used to boost the teacher's robustness and brings a robust accuracy of 54\% against AA. The results are illustrated in Table \ref{tab:AWPTeacher}.
It shows that the \textbf{best} checkpoint of student model has 0.49\% and 0.13\% improvement in natural and robust accuracy, respectively, and the \textbf{last} checkpoint slightly degrades in robustness but gains 0.7\% improvement in natural accuracy. 
It can be concluded that RSLAD can indeed boost the small models' robustness with various kinds teacher models.

\begin{table*}[h]
\begin{center}
\setlength{\tabcolsep}{2mm}{
\caption{White-box robustness results of ResNet-18 on CIFAR-10. The best results are \textbf{boldfaced}. $^{*}$ indicates the teacher is SAT+AWP. }
\label{tab:AWPTeacher}
\scalebox{0.95}{
\begin{tabular}{c|ccc|ccc}
\toprule
\multirow{2}{*}{Method} & \multicolumn{3}{c|}{\textbf{Best Checkpoint}} & \multicolumn{3}{c}{\textbf{Last Checkpoint}} \\
 & Clean& PGD$_{\textup{TRADES}}$ & AA & Clean& PGD$_{\textup{TRADES}}$ & AA\\
\midrule
Natural& 94.65\%& 0.0\%& 0.0\% & 94.65\%& 0.0\%& 0.0\% \\
\hline
SAT& 83.38\%& 51.11\%& 45.83\% & 84.44\%& 48.72\%& 43.64\% \\
\hline
TRADES & 81.93\% & 53.68\%&49.23\% & 82.20\% &53.66\%& 49.27\% \\ 
\hline
RSLAD & 83.38\% & 55.94\%&51.49\% & 83.33\% &55.61\%& \textbf{51.32\%} \\
RSLAD$^{*}$ & 83.87\% & \textbf{56.78\%} & \textbf{51.62\%} & 84.03\%&\textbf{56.52\%}&51.09\% \\
\bottomrule
\end{tabular}
}}
\end{center}
\end{table*}

\end{document}